\documentclass[superscriptaddress,onecolumn,preprint]{revtex4-1}
\usepackage{graphicx, epsfig}
\usepackage{overpic,color}
\usepackage{textcomp}
\usepackage{setspace}
\usepackage{amssymb,amsmath} 
\usepackage{tabulary}
\usepackage{multirow}
\usepackage{enumerate}
\usepackage[usenames,dvipsnames,svgnames]{xcolor}

\begin{document}

\title {Particle-Level Modeling of the Charge-Discharge Behavior of Nanoparticulate Phase-Separating Li-Ion Battery Electrodes}
\author{Bernardo Orvananos}
\affiliation{Department of Materials Science and Engineering, University of Michigan}
\author{Todd R. Ferguson}
\affiliation{Department of Chemical Engineering, Massachusetts Institute of Technology}
\author{Hui-Chia Yu}
\affiliation{Department of Materials Science and Engineering, University of Michigan}
\author{Martin Z. Bazant}
\affiliation{Department of Chemical Engineering, Massachusetts Institute of Technology}
\affiliation{Department of Mathematics, Massachusetts Institute of Technology}
\author{Katsuyo Thornton}
\affiliation{Department of Materials Science and Engineering, University of Michigan}
\date{\today}

\begin{abstract} In nanoparticulate phase-separating electrodes, phase separation inside the particles can be hindered during their charge/discharge cycles even when a thermodynamic driving force for phase separation exists. In such cases, particles may (de)lithiate discretely in a process referred to as mosaic instability. This instability could be the key to elucidating the complex charge/discharge dynamics in nanoparticulate phase-separating electrodes. In this paper, the dynamics of the mosaic instability is studied using Smoothed Boundary Method simulations at the particle level, where the concentration and electrostatic potential fields are spatially resolved around individual particles. Two sets of configurations consisting of spherical particles with an identical radius are employed to study the instability in detail.  The effect of an activity-dependent exchange current density on the mosaic instability, which leads to asymmetric charge/discharge, is also studied. While we show that our model reproduces the results of  a porous-electrode model for the simple setup studied here, it is a powerful framework with the capability to predict the detailed dynamics in three-dimensional complex electrodes and provides further insights into the complex dynamics that result from the coupling of electrochemistry, thermodynamics, and transport kinetics. 
\end{abstract}

\maketitle

\section {Introduction}

Many electrode chemistries undergo phase transformation during their charge-discharge cycle. Examples of these materials include graphite, some transition metal oxides, and olivine phosphates \cite{ferguson2013}. When the particles of these electrode materials are sufficiently large, the particles tend to phase separate into their different stable phases generating interfaces. However, when the particles are nanosized, phase separation may be hindered. The dynamics in such cases is still poorly understood. As nanostructured materials are employed in many applications, a better understanding of their phase separation behavior is of critical importance. 

The interest in phase-separating electrode materials has been driven mostly by the promising performance of nanoparticulate lithium iron phosphate, LiFePO$_4$ (LFP). LFP, first explored by Padhi {\it et al.} as an electrode application in 1997 \cite{Padhi:1997dq}, was initially not considered a suitable electrode material upon its discovery because of its low ionic diffusivity and electronic conductivity. Although this difficulty has been overcome by nanosizing the particles \cite{Huang:2001hh}, carbon coating \cite{Wang:2008ix}, and doping with other metallic cations \cite{Chung:2002bi}, it remains unclear whether phase separation occurs and how the phase separation proceeds in nanoparticulate LFP. To answer this question, a variety of models have been proposed. Several models have proposed different paths for phase separation  \cite{Srinivasan:2004hga,Laffont:2006dr,Tang:2009dj,singh2008,Delmas:2008cb}, while others have proposed that the particles remain homogeneous without nucleation \cite{Malik:2011xd}. Bai {\it et al.}\ proposed that phase separation could occur at low currents while large enough currents can suppress phase transformation \cite{Bai:2011je}, thus hindering the emergence of distinct phases during lithiation/delithiation. Cogswell and Bazant further developed on this model \cite{Cogswell:2012er}, and recently suggested that a wetting nucleation process can facilitate phase separation in nanoparticles \cite{Cogswell:2013}.

On the electrode scale, where $10^{10}-10^{17}$ of electrode particles are present \cite{Dreyer:2010ic}, inhomogeneities in the electrode potential and the difference in the particle sizes can lead to large current density variations in space. For example, smaller particles, which have larger surface area per volume ratios, tend to fill preferentially. This can lead to large local current densities for those particles, which can suppress phase separation even at low cell currents \cite{Ferguson:2012vk,ferguson2013,Bai:2011je}. As a result, a different type of phenomenon, referred to as the ``mosaic instability'' \cite{Burch:2009wg}, takes place. In this case, ``phase separation" occurs between particles instead of within each single particle, and thus nanoparticles tend to be either nearly fully lithiated or fully delithiated.

In this paper, we investigate the dynamics of interparticle phase separation ({\it i.e.}, mosaic instability) during lithiation and delithiation (discharge and charge) in phase separating electrodes consisting of nanoparticles. To focus on the interactions between particles, we constrain ourselves to the case where all the particles are of the same size. This eliminates the effect of particle size distribution and allows us to focus on the effect of the location of the particles within the electrode. We model the nanoparticles as a Li solid solution based on the theory mentioned above. To elucidate the interactions between particles, we first analyze a cell containing two nanoparticles (Sec.~\ref{sec:2part}). The simulations are then extended to a larger domain containing 26 nanoparticles in a unit cell that is periodic in the direction normal to the cell current (Sec.~\ref{sec:const_j0}-\ref{sec:var_j0}). The interaction dynamics under varying applied currents is investigated. Furthermore, the effect of the exchange current density is studied by taking the exchange current density to be independent of the activity of the nanoparticles (Sec.~\ref{sec:const_j0}) and by using a cathode-activity-dependent exchange current density (Sec.~\ref{sec:var_j0}). Analyses are performed to determine what critical concentration difference is required to trigger the instability, what controls the size of the group of particles that undergo phase transformation simultaneously in each instability, and how two mechanisms (Li redistribution and constant (de)lithiation) compete during the instability (Sec.~\ref{sec:analysis}). Finally, we compare our simulations with that of a macro-homogeneous model based on the porous electrode theory as a verification of our results (Sec.~\ref{PET}). 

Our simulations show that the interparticle interactions are more significant at lower applied currents, which lead to sudden rises and drops in the voltage profile. The resulting dynamics is similar to the mosaic instability, but instead of being governed by the particle size difference, it is by the location of the particles in our case. Note that in actual cells, the mosaic instability would occur as a combination of both location and size. At a sufficiently high current, the interaction is suppressed. Our work reveals the detailed electrochemical dynamics in an individual particle level, and extends to the collective behavior of many particles as an electrode. Although this work may explain the charge/discharge behavior of nanoparticulate LFP, the insights gained for mosaic instabilities can be applied to any physical system in which its tendency of phase separation is suppressed.

\section {Model}
\subsection{Overview}
We present a model for the dynamics of the lithiation-delithiation process at the particle-level based on the assumption that the nanoparticles considered are too small to exhibit phase separation within a particle. Hereafter, a nanoparticle that is assumed to exhibit solid-solution behavior is simply referred to as a ``particle" throughout this paper, and the term ``mosaic instability," originally defined as preferential particle filling resulting from the different particle sizes, is expanded to include a group-by-group filling.

In the cathode particles, the transport of lithium is modeled using Fick's diffusion law, because it is assumed that phase separation is inhibited, as described above. It is assumed that the cathode particles are electrically well connected to the current collector through a carbon black network, and consequently the potential of electrons in the electrode is assumed to be uniform throughout the entire cathode.  The transport of lithium ions in the electrolyte is driven by the gradients of concentration and electrostatic potential, which correspond to diffusion and migration, respectively; they are modeled here using the Nernst-Planck equation. Electroneutrality is assumed throughout the electrolyte, and thus the potential field in the electrolyte is calculated according to the charge conservation condition. The reaction of Li-ions with electrons is modeled using a modified form of the Butler-Volmer Equation \cite{Ferguson:2012vk,Bazant:2013ik}.
 
The Smoothed Boundary Method (SBM) is used to numerically implement the electrochemical model of the cell because of its advantages in handling problems with multiple physics and multiple phases. This method was first proposed by Bueno-Orovio {\it et al.} \cite{BuenoOrovio:2006fi,BuenoOrovio:2006gy,BUENOOROVIO:2006cm}, and was further extended and generalized by Yu {\it et al.} \cite{Yu:2012oq}. In this method, a continuous domain parameter, $\psi(x)$, is used to distinguish the different domains (phases), {\it i.e.}, the electrolyte and cathode in this case.  The domain boundary is defined as a region of finite thickness, rather than a sharp boundary, from which the name, the Smoothed Boundary Method, originates.  The cathode particles are defined as the regions where $\psi =1$ and the electrolyte where $\psi =0$. The value of the domain parameter smoothly transitions from zero to one over the region near the domain boundary between the electrolyte and the cathode particle, giving a finite thickness to the interfacial region (where $0<\psi<1$). The original governing equations that describe lithium transport in the electrolyte and cathode particles and the charge conservation in the electrolyte are reformulated into the SBM form. The details of the mathematical derivation of the equations can be found in Ref.\ \cite{Yu:2012oq}. In the next subsections, we present the details of the model and introduce its governing equations in both the original and their SBM forms.

\subsection {Transport in the nanoparticles}
\label {thermo}
Here, we present the equations for the transport of Li in the particles and the underlying thermodynamics. First, we introduce the free energy and chemical potential that govern the thermodynamics for Li transport in the particles. We use the regular solution model, in which both the entropic and enthalpic effects are included. The free energy of Li in the particles, $f$, takes the form of
\begin {equation}
f = RT \bigg [ X_s \ln (X_s) + (1-X_s) \ln (1-X_s)+\Omega X_s (1-X_s ) \bigg ],
\label{f_RSM_1}
\end {equation} 
where $X_s$ is the site fraction of Li in $ \rm {Li_{\it x}FePO_4}$, $R$ is the gas constant, $T$ is the absolute temperature and $\Omega$ is the regular solution parameter. The subscript $s$ denotes ``solid.'' The value of $X_s$ is related to the Li concentration, $C_s$, by $X_s = C_s / \rho_{Li}$, where $\rho_{Li}$ is the Li concentration when $x=1$. The chemical potential, $\mu_s$, is defined as the derivative of the free energy with respect to the Li site fraction:
\begin {equation}
\mu_s = \frac {\partial f} {\partial X_s} =  RT \bigg [ \ln \bigg ( \frac { X_s} {1-X_s} \bigg ) +  \Omega (1-2 X_s)  \bigg ].
\label{mu_RSM_1}
\end {equation} 
When $\Omega$ has a value greater than two, the free energy becomes a double-well function and the chemical potential becomes non-monotonic. This represents a state where phase separation is thermodynamically favored. The range of composition in which the system is unstable with respect to composition fluctuations is referred to as the spinodal region; and the concentration boundaries of the region are referred to as the spinodal points. They correspond to the site fraction at which the chemical potential is at the local minimum or local maximum (or the inflection points in the free energy).

We now introduce the governing equations for the transport of Li in the particles. To model concentration evolution in systems that tend to phase separate, the Cahn-Hilliard equation \cite{CAHN:1958uq}, the phase field equation with conserved order parameter, is commonly used. The Cahn-Hilliard equation is given by
\begin {equation}
\label {CH}
\frac {\partial C_s} {\partial t} = \nabla \cdot \bigg ( M_s \nabla ( \mu_s - \kappa \nabla^2 C_s ) \bigg ) \in V_s,
\end {equation}
where $t$ is time, $M_s$ is the mobility of Li in the particles, $\kappa$ is energy gradient penalty, and $V_s$ represents the bulk of the cathode particles. The term $\kappa\nabla^2 C_s$ accounts for the penalty of having an interface within the particle.
 
As mentioned earlier, we assume phase separation cannot occur within the individual particles. This can be justified as follows. At low currents, the non-monotonic equilibrium potential produces the mosaic instability at the electrode scale, causing large local current densities. These large local current densities suppress intraparticle phase transformation \cite{Bai:2011je}. At higher currents, the current density is more homogeneous throughout the electrode, but still the higher overall current suppresses the phase transformation. Also, the diffusion time of Li in the particles is much shorter than both the transport time of Li in the electrolyte and the reaction rate. This allows us to model the particles near the ``pseudocapacitor limit,'' at which particles are assumed to be homogeneous. Therefore, the detailed Li diffusion model inside the individual particles would not be the primary concern of this paper. For simplicity and numerical efficiency, we model the concentration evolution in the particle with Fick's law of diffusion, rather than the Cahn-Hilliard equation (Eq.\ \ref{CH}).  Consequently, we evolve. 
\begin {subequations}
\begin {equation}
\label {SimpleDiff}
\frac {\partial C_s} {\partial t} = \nabla \cdot D_s \nabla C_s \in V_s,
\end {equation}
where $D_s$ is the diffusion coefficient of Li in the particles. For the cases examined in this work, the concentration of Li in the particles is nearly uniform, but we use this model to retain generality. As in the case for electrolyte, the boundary condition at the particle-electrolyte interface is set by the reaction rate, accounting for Li (de)intercalation:  
\begin {equation}
\label {SimpleDiff2}
\vec n \cdot \mathbf{J} = r_{Li} \in A,
\end {equation}
\end {subequations}
where $r_{Li}$ is the reaction rate, $\vec n$ represents the normal unit vector in direction from the electrolyte to the particle, and $\mathbf {J}$ is the flux at the particle-electrolyte interfacial region, $A$. Equations \eqref{SimpleDiff} and \eqref{SimpleDiff2} are reformulated to their SBM form consisting of bulk and boundary terms \cite{Yu:2012oq},
\begin {equation}
\label {SBM_Cs3}
\frac {\partial C_s} {\partial t} = \frac {1} {\psi} [ \nabla \cdot \psi D_s \nabla C_s] + \frac { | \nabla \psi | } {\psi} r_{Li}. 
\end {equation}
Note that Eq.\ \eqref{SBM_Cs3} reduces to the original equation in Eq.\ \eqref{SimpleDiff} in the bulk of the particles, where $\psi =1 $.   
  
\subsection {Transport and charge conservation in the electrolyte} 
 The governing equations for the transport of the salt and the charge conservation in the electrolyte are now presented. We adopt the dilute solution model for salt diffusion in a binary electrolyte where one cationic and one anionic species are present in a neutral solvent, as described in the textbook of Newman and Thomas-Alyea \cite{Newman:2004uq}. Thus, the governing equation for the dynamics of the salt in the electrolyte is
 \begin {subequations} 
\begin {equation} 
\label {ef_diff2}
\frac { \partial C_l} { \partial t} = \nabla \cdot (D_{amb} \nabla C_l) \in V_l, 
\end {equation}  
where $C_l$ is the concentration of the salt in the electrolyte, $t$ is time, $D_{amb}$ is the ambipolar diffusion coefficient and $V_l$ represents the bulk of the electrolyte domain. The subscript $l$ denotes ``liquid.'' The reaction rate accounting for the transfer of Li ions to and from the electrodes is similar to that used by Ferguson and Bazant \cite{Ferguson:2012vk}, which assumes bulk neutrality in a binary electrolyte with constant ion diffusivities,
\begin {equation} 
\label {ef_diff2b}
\vec n \cdot \mathbf{J} = - (1-t_+) r_{Li} \in A , 
\end {equation}  
\end {subequations}
where $t_+$ is the transference number of the cation. The ambipolar diffusion coefficient is an effective diffusion coefficient accounting for both diffusion and migration of the salt and is defined as: $ D_{amb}= D_+ D_- (z_+ - z_-)/(z_+D_+ - z_-D_-)$, where $D_+$ and $D_-$ are the diffusion coefficient of the cation and anion respectively, and $z_i$ is the charge number of $i$th species. We assume that $ D_{amb}$ is constant. We use the SBM for the simulation, where Eqs.~\eqref {ef_diff2} and \eqref {ef_diff2b} are combined using the SBM domain parameter $\psi$ to obtain
\begin {equation}
\label {Cs_time3}
 \frac {\partial C_l} {\partial t} = \frac {D_{amb}} {1-\psi} \bigg [ \nabla \cdot \bigg ( (1-\psi) \nabla C_l \bigg ) \bigg ] - (1-t_+) \frac { | \nabla \psi | } {1- \psi} r_{Li}.
\end {equation}
Note that Eq.~(\ref{Cs_time3}) reduces to the original equation in Eq.~(\ref{ef_diff2}) in the electrolyte, where $1-\psi =1 $. At the particle-electrolyte interface where $|\nabla \psi| \neq 0$, the reaction rate serves as the flux boundary condition. The regions where $1-\psi =0$ are outside of the domain, and the solutions in these regions are nonphysical and irrelevant.

Here, we neglect the effect of the double layers because, for the range of the current we consider, depletion in the electrolyte is limited and concentrations remain high ($\sim$1M), leading to thin double layers. (For a model of porous electrodes with double layers, see Ref.\ \cite{biesheuvel2012}.) Since the electrolyte is assumed to be electroneutral, the concentrations and the total amounts of each ionic species (anions or cations) in the electrolyte are equal. The anions are considered to be inert so that the total amounts of both species in the electrolyte must remain constant at all times.
As a result, as the Li reacts at the cathode particle surfaces, we assume that the corresponding amount of lithium is reacting at the anode-electrolyte interface. 

The transport of a charged species through the electrolyte results in an ionic current, which can be decomposed into the terms corresponding to diffusion and migration.  The current density vector, which accounts for both diffusion and migration, $\mathbf{i}$ is given by \cite{Newman:2004uq}
\begin {equation}
\label {curr}
 \mathbf{i}  = -z_+ \upsilon_+ F \bigg[ \frac {F} {RT} (z_+D_+ - z_-D_-) C_l \nabla \phi_l + (D_+ - D_- ) \nabla C_l \bigg],
\end {equation}
where $\upsilon_{+}$ is the number of cations produced by the dissociation of the salt, $F$ is Faraday's constant, and $\phi_l$ is the electrostatic potential of the electrolyte.

According to the charge conservation condition, current is conserved in the absence of a source or sink (or equivalently, in the absence of reaction). Therefore, the divergence of the current equals zero in the electrolyte away from the interfacial regions:
\begin {subequations}
\begin {equation}
\label {curr_cons_l}
 - \frac { \nabla \cdot \mathbf{i} } {z_+ \upsilon_+ F}=\nabla \cdot \bigg [ \frac {F} {RT} (z_+D_+ - z_-D_-) C_l \nabla \phi_l \bigg ] + \nabla \cdot [ (D_+ - D_- ) \nabla C_l] =  0 \in V_l.
\end {equation}
On the other hand, the divergence of the current is proportional to the reaction rate at the interfaces:
\begin {equation}
\label {curr_cons_l2}
 - \frac { \nabla \cdot \mathbf{i} } {z_+ \upsilon_+ F} = \frac {r_{Li}} {v_+} \in A.
\end {equation}
\end {subequations}
These equations can be combined into a single equation in the SBM formulation, which becomes
\begin {equation}
\label {SBM_phil}
\nabla \cdot [( 1- \psi) \frac {F} {RT} (z_+D_+ - z_-D_-) C_l \nabla \phi_l ] = | \nabla \psi | \frac {r_{Li}} {v_+} + \nabla \cdot [(1-\psi) (D_- - D_+ ) \nabla C_l]. 
\end {equation}
This equation is solved to obtain the electrostatic potential in the electrolyte with the flux boundary condition imposed at the particle-electrolyte interfaces.  The electrostatic potential enters into the Butler-Volmer Equation to determine the reaction rate, as described below.

\subsection {Reaction at the particle-electrolyte interface} 
\label {Sec:rxn}  

Now, we introduce the interfacial kinetics model of the redox reaction: $\rm {Li^+ +e^-  + FePO_4}$ $ \rm {\rightleftarrows LiFePO_4}$ for the case of LFP. In the forward reaction, the Li ions (Li$^+$), dissolved in the electrolyte, react with the electrons from the electrode, to produce neutral Li atoms that are inserted into the cathode particles.  In the reverse reaction, the neutral Li in the cathode loses an electron and is extracted into the electrolyte.  We assume that the reaction takes place at the particle-electrolyte interface, and that the transport of electrons is sufficiently fast so that they are abundantly available for the reaction. 

The reaction rate is modeled using the modified Butler-Volmer equation proposed in Refs.\ \cite{Bazant:2013ik,Bai:2011je}, 
\begin {equation}
\label {BV_i0}
r_{Li}=r_{Li}^{BV}= \frac {i_0}{F} \bigg [ \exp \bigg ( -\frac { \alpha  F} {RT} \eta \bigg ) - \exp \bigg (   \frac { (1- \alpha ) F} {RT} \eta \bigg ) \bigg ],
\end {equation}
where $i_0$ is the exchange current density, defined as a function of Li activity in the cathode (unlike the standard Butler-Volmer Equation, which is linearly dependent on the concentration), $\alpha$ is the transfer coefficient, and $\eta$ is the overpotential. 
The overpotential, $\eta$, is defined as $(\phi_s - \phi_l) - \phi_{Eq}$, where 
$(\phi_s - \phi_l)$ is the potential difference across the particle-electrolyte interface, and $\phi_{Eq}$ is the equilibrium potential. The value of $\phi_{Eq}$ is approximated by Nernst equation assuming Li metal as a reference, $\phi_{Eq} = V_{OC} - \mu_s / F$, where, $V_{OC}$ is the plateau value of the open circuit voltage (OCV).

As mentioned above, in the modified Butler-Volmer equation, the exchange current density is a function of the activities of Li in the electrolyte and the particles, which leads to nonlinear dependence on the concentration of Li, unlike in the standard Butler-Volmer equation. In a dilute electrolyte, the activity can be approximated by the normalized concentration, $a_l=C_l / C_l^0$, where $C_l^0$ is the concentration at which $i_0$ was measured. In the particles, the activity is defined by the Arrhenius equation of the chemical potential as $a_s= \exp (  \mu_s / R T )$. Thus, the exchange current density is given by
\begin {equation}
i_0= \frac {F (k_c^0a_l)^{1-\alpha}(k_a^0a_s)^{\alpha}} {\gamma_{TS}} = i_0' \sqrt {  a_l X_s(1-X_s) \exp [ \Omega (1 - 2 X_s)] }. 
\label {i_0}
\end {equation}
Here, it has been assumed that $\alpha = 0.5$ and $ i_0' = F (k_c^0)^{1-\alpha}(k_a^0)^{\alpha}$, where $i_0'$ is the exchange current coefficient, $k_c^0$ and $k_a^0$ are the standard rate constants of cathodic and anodic reactions, respectively, and $\gamma_{TS} $ is the chemical activity coefficient of the transition state approximated as $(1-X_s)^{-1}$ to account for the site availability \cite{Bai:2011je}.
 
\subsection {Simulation configuration}
We consider a cell that contains a cathode consisting of equal-sized nanoparticles immersed in a LiPF$_6$ electrolyte, a separator represented by empty space filled with electrolyte, and lithium metal foil as the anode, for which the Li concentration and chemical potential are constant. A square-prismatic computational box is used in the simulations, which spans 1152 nm in the direction from the anode to the cathode current collector ($z$-axis) and 64 nm in other directions ($x$- and $y$-axes). The anode-electrolyte interface is located at the boundary at $z=0$, and the cathode current collector at $z=1152$ nm. Periodic boundary conditions are imposed on the $x$-$z$ and $y$-$z$ planes of the computational box. Thus, this configuration represents a planar cell, of which in-plane dimensions (in $x$-$y$ plane) are much larger than the depth (in $z$-direction) of the cell. Such an arrangement is convenient for investigating the effect of particle locations on the mosaic instability in the depth direction. 

A Cartesian grid system with uniform grid spacing of 2 nm in the computational box was used. A second-order central finite difference scheme in space with a first-order Euler explicit time scheme was employed to solve the Li transport in the particles. An alternative-direction-line-relaxation (ADLR) solver \cite{Velde-E-F:1994fk,Hofhaus:1996uq,Yu:2012oq} was applied to solve for the electrostatic potential and the Li transport in the electrolyte.

\begin{figure}[htp] 
\begin{center}
\includegraphics[width=\textwidth]{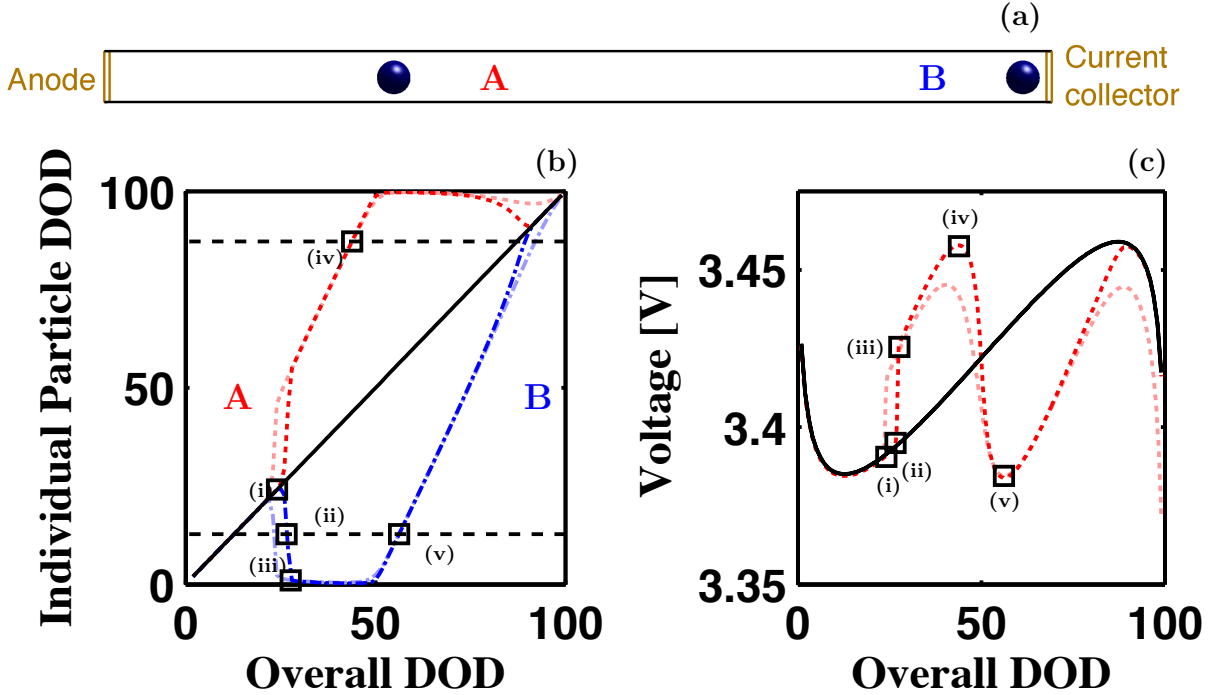}
\end{center}
\caption{(a) Two particle configuration used in these simulations. (b) DOD of the individual particles with respect to the overall DOD. The dashed red lines correspond to the DOD of the particle closer to the separator (particle A), the dash-dotted blue lines to the DOD of the particle closer to the cathode current collector (particle B) and the black line to the overall DOD.  (c) Simulated voltage during the lithiation process with respect to the overall DOD. The red lines represent the obtained voltage from the simulations and the black line corresponds to the equilibrium potential. In (b) and (c), the darker lines represent the case when $i_0$ is assumed to be activity independent and the lighter lines represent when this dependence is included. Labels (i)-(v) highlight the key points of the lithiation process. (Colors are only shown in the online version)}
\label{insert2p}
\end{figure}

Simulations with two different configurations were conducted: one with two particles and the other with 26 particles in the cell. In the first set, two particles 40 nm in diameter were located 728 nm apart (measured between the closest surfaces), one particle was 332 nm from the anode and the other was 12 nm from the cathode current collector with respect to the nearest surfaces; see Fig. \ref{insert2p}(a). This two-particle simulation is designed to illustrate the cell voltage response during the interaction between particles. In the second simulation configuration, the cathode contains 26 particles in a body-centered-cubic (BCC) arrangement, where the shortest distance between particle surfaces is 15 nm; see Fig.\ \ref{insert}(a). The cathode spans 852 nm in the $z$-direction, while the separator (region of electrolyte without particles) has a length of 300 nm. The cathode region of this cell has a volume fraction of approximately 25.3\% for the active particles.  While this particle volume fraction is smaller than an actual battery cathode, it allows us to elucidate the behavior of interacting particles, and we expect the qualitative findings to remain valid for the range of current we examine in this work.  This set of simulations with this given arrangement provides information for the many-particle dynamics. The structure of the cathode was defined by the domain parameter in the SBM as previously described. A constant applied current was maintained during the simulation (of either the lithiation or delithiation process, depending on the simulation) by adjusting the value of the electrostatic potential of the particles. Note, however, that the imposed value of the applied current was allowed to fluctuate $\pm 1.5\% $ to facilitate the faster convergence of the ADLR scheme and resulting improved numerical efficiency. 

We choose the average applied current density (normalized to the total particle surface), $\overline i$, to be one of the controlling parameters, 
\begin {equation}
\overline i =  \frac { \sum\limits_{j=1}^{N_{tot}}  \int\limits_{S_j} r_{Li,j}F dS_j} { \sum\limits_{j=1}^{N_{tot}}  \int\limits_{S_j}  dS_j}  
\end {equation}
where $S_j$ and $r_{Li,j}$ are the surface and reaction rate of particle $j$, respectively, and $N_{tot}$ is the total number of particles.  Throughout the text, $\overline i$ is specified as a fraction of the exchange current coefficient, $i_0'$. We do so to emphasize that the observed dynamics at a given rate depends on the ratio $\overline i /i_0'$ and that there is a certain degree of uncertainty in the literature values of $i_0'$. Note that a constant current density with respect to the particle surface is equivalent to a constant C-rate since the particle size remains unchanged. However, the current density scales with the particle surface while the C-rate, scales with the particle volume (or mass). Consequently, for a system with a given mass under the same C-rate but with larger particles, a larger current density at the surface of the particles will be observed.

The physical parameters used in the simulations are given as follows. For the cathode, the diffusion coefficient of Li in the particles is $1\times 10^{-12}$ cm$^2$/s \cite{Zhu:2010kf}, assumed isotropic for simplicity, and the interstitial site density is 0.0228 mol/cm$^3$.  In the electrolyte, the diffusion coefficients for the cation (Li$^+$) and anion (PF$_6^-$) are $1.25 \times 10^{-6}$ cm$^2$/s and $4.0 \times 10^{-6}$ cm$^2$/s, respectively, to match the work of Ferguson and Bazant \cite{Ferguson:2012vk}, which were based on experimental results \cite{Capiglia:1999uq,Valoen:2005kx}. The electrolyte has an average concentration of 1 M (consistent with the molarity at which the ionic diffusivities and the exchange current density were measured). The dissociation number of the cation is equal to one. For the modified Butler-Volmer equation, we take $i_0'=1.75 \times 10^{-6}$ A/cm$^2$. Here, we have scaled an experimental value of $i_0'$ similar to the one from Refs.\ \cite{Pasquali:2008fk,Zhu:2011io} by 1/100, as was done in Ref.\ \cite{Bai:2011je}, since the experimental value was measured per the macroscopic cross-sectional area of the cathode, not the actual particle surface area, which is needed for our simulation. The OCV plateau of LFP is taken to be 3.422 V \cite{Dreyer:2010ic} and the regular solution parameter, $\Omega$, is set to 4.5 \cite{Cogswell:2012er} for evaluating the chemical potential. This regular solution parameter gives a non-monotonic voltage profile with a difference between the local minimum and local maximum of approximately 74 mV; see the equilibrium potential profile in Fig.\ \ref {insert2p}(c).   
  
\section{Results: dynamics during lithiation in a two-particle cell}
\label{sec:2part}
 
We first conducted the simulation of a two-particle cell to illustrate the interparticle interactions and the corresponding voltage response, in a similar way as in Ref.\ \cite{Burch:2009wg}. The cell was lithiated at an applied current density, $\overline i $, which is chosen to be 2\% of $ i_0'$ to examine the low current regime. This loading condition is approximately a C/12 rate. In the first case, we focus exclusively on the interparticle interactions. We first ignore the dependence of the exchange current density on the activity of the cathode particles and $i_0$ is assumed to only depend on the electrolyte activity, $i_0 = i_0' \sqrt {a_l}$. Hereafter, we refer to the cathode-activity simply as ``activity" for convenience.

The depth of discharge (DOD) for the two individual particles and the voltage profile during the lithiation process are shown in Figs.\ \ref{insert2p}(b) and (c), respectively. Here, we refer to the particle close to the separator as particle A and the one close to the cathode current collector as particle B. Since the cell is lithiated under a constant current, the average DOD increases linearly. However the DOD of the individual particles exhibits different behavior. Five points, (i) through (v), on the curves are noted to illustrate the unique dynamics of the interparticle interactions.  In the early stage of discharge (prior to point (i)), both of the particles lithiate at a similar rate. However, particle A lithiates slightly faster because both the ion concentration and electrostatic potential in the surrounding electrolyte are higher than that of particle B. Thus, particle A reaches the concentration level of the lower spinodal point first and undergoes rapid lithiation (see point (i)), during which lithium is extracted from particle B (between point (i) and (ii)). Once the concentration level of particle B drops below that of the lower spinodal point, a sudden rise in the voltage curve is observed; see point (ii) in Figs.\ \ref{insert2p}(b) and (c). After particle B fully delithiates, the slope of the voltage curve decreases due to the slow lithiation of particle A because it can no longer extract lithium from particle B; see point (iii). When particle A reaches the higher spinodal point, the voltage decreases; see point (iv). In the meantime, the concentration level of particle B slowly increases. After particle B reaches the lower spinodal point, the voltage rises again; see point (v). During the lithiation of particle B, lithium is extracted from particle A to a lesser degree; see the DOD of particle A between point (v) and the end of the evolution.

The dynamics observed above can be understood as follows. To maintain a constant current, the applied voltage is adjusted to a specific value. When such voltage resides in the gap between the local minimum and local maximum of the equilibrium potential curve (referred to hereafter as the ``voltage window'') and falls in-between the equilibrium potentials of the two particles, one particle is driven to lithiate and the other to delithiate, producing a mosaic instability. 

\begin{figure}[htp] 
\begin{center}
\includegraphics[width=.5\textwidth]{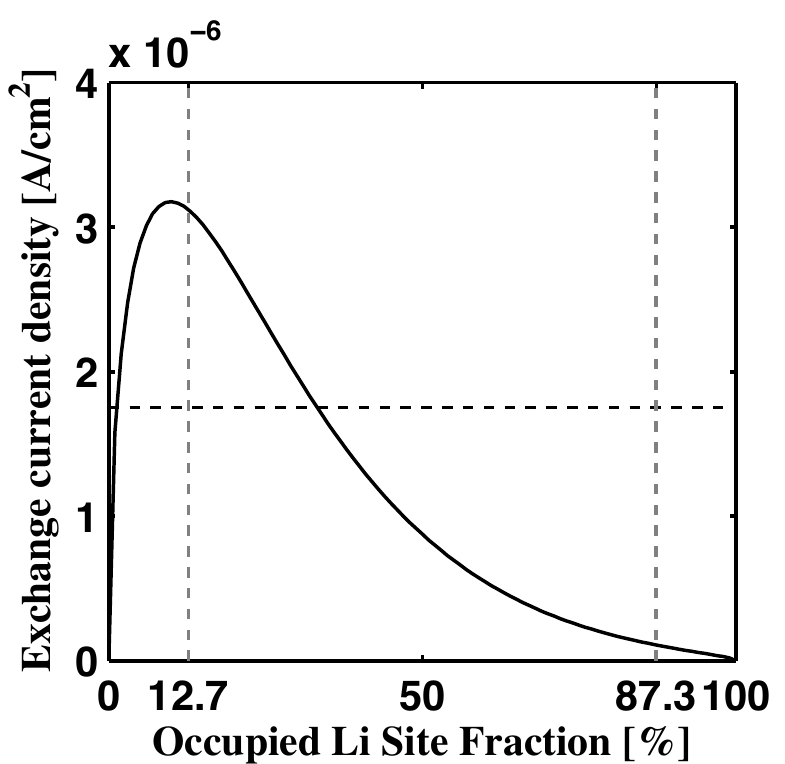}
\end{center}
\caption{Exchange current density. Solid line: $i_0$ as a function of the occupied Li site fraction as expressed in Eq.~\eqref {i_0}. Dashed horizontal line: value of $i_0$ independent of the Li site fraction. The maximum value of the function is located near the lower spinodal point. The value of the exchange current density at the lower spinodal point is 28.6 times higher than that at the larger spinodal point. The vertical dashed lines represent the spinodal points.   }
\label{i0}
\end{figure}

Next, we conduct the simulation where the exchange current density is a function of the activity of the particles, as given in Eq.~\eqref{i_0}; see Fig.\ \ref{i0}. The form of the exchange current density results from the regular solution model, as explained in Sec.~\ref{Sec:rxn}. It has the maximum value when the Li concentration is close to the lower spinodal point, as this is where both the chemical potential and the site availability are large.  The difference is substantial -- the exchange current density is 28.6 times larger at the lower spinodal point than at the higher spinodal point. Therefore, the magnitude of the required overpotential to maintain a constant current density is smaller at low concentrations and is larger at high concentrations. It is important to note that this exchange current density is based on the regular solution model.  For higher accuracy, a more realistic exchange current density is required.  

The simulation results are given in the lighter lines in Figs.\ \ref{insert2p}(b) and (c), and show three features that differ from the case with exchange current density independent of the activity. First, the mosaic instability occurs at a lower concentration, since the magnitude of the overpotential at low concentrations is smaller (this will be further explained in Sec.~\ref{CCD}). Second, the peak of the voltage rise is smaller (Fig.\ \ref{insert2p}(c)), since the overpotential at higher concentrations is larger. Last, at the very end of the process, the exchange of Li between particle A and particle B is almost completely suppressed (Fig.\ \ref{insert2p}(b)). This is caused by the much smaller exchange current density of the particles at that concentration.
  
\section {Results: lithiation and delithiation with exchange current density independent of the cathode activity}
\label {sec:const_j0}
 
\begin{figure}[htp] 
\begin{center}
\includegraphics[width=.75\textwidth]{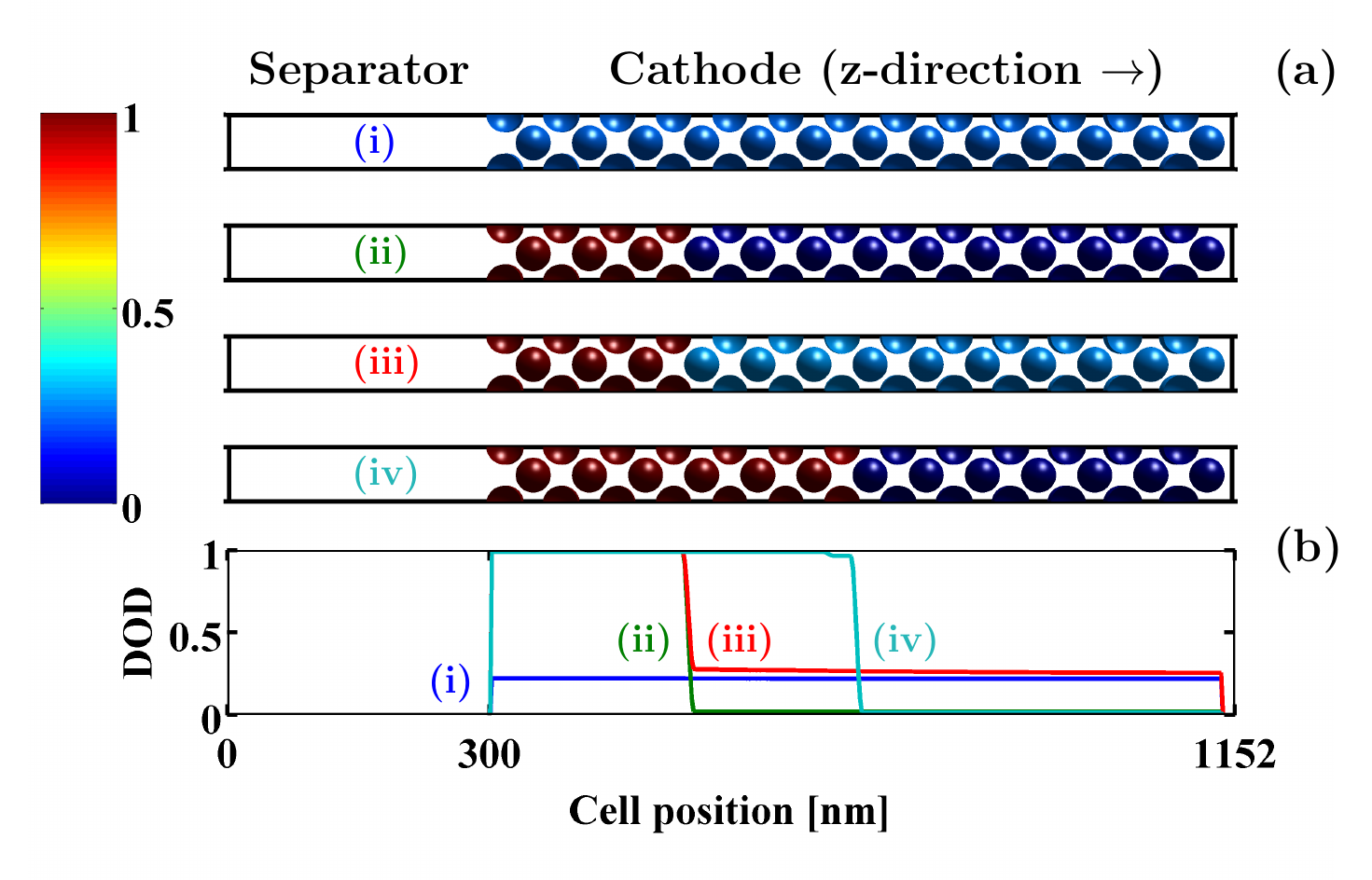}
\end{center}
\caption{Dynamics observed during lithiation when $i_0$ is assumed independent of the activity of the cathode. (a) Surface Li concentration at four different times. (i) All particles lithiate in a nearly uniform manner (DOD = 22\%). (ii) The particles closer to the separator transform to Li-rich phase, and the particles farther away return to a Li-poor phase (DOD = 28\%). (iii) The Li-poor particles lithiate in a nearly uniform manner (DOD = 46\%). (iv) Another group of particles transforms to the Li-rich phase, delithiating the remaining particles (DOD = 50\%). Note that, because the process is reaction limited, each particle has nearly constant concentration ({\it i.e.}, the surface concentration is approximately equal to the bulk concentration) (b) DOD averaged over the particle region in a cross section in the x-y plane.}
\label{insert}
\end{figure}

Having developed an understanding of how two particles can interact, we now examine the dynamics with a larger number of particles via a simulation of the 26-particle configuration mentioned earlier. As in the two-particle configuration, the simulation was also performed at $\overline i  = 2\%~i_0'$ (C/12 rate) and the dependence of the exchange current density on the activity of the particles was ignored by assuming $i_0 = i_0' \sqrt {a_l}$ in this section. Figure \ref {insert} shows Li concentration in the particles at four different times during lithiation. First, all the particles lithiate in a fairly even manner up to the cell DOD of 22\%; see Fig.\ \ref {insert}(i). Next, a group of particles close to the separator simultaneously undergo fast lithiation, extracting lithium from the rest of the particles. As a result, a mosaic instability occurs at the cell DOD of 28\%, where seven layers of particles are nearly fully lithiated, while the remaining 19 layers are nearly fully delithiated; see Fig.\ \ref {insert}(ii). The process is repeated with the remaining delithiated particles.  They lithiate in a fairly uniform manner (see Fig.\ \ref {insert}(iii)), and then undergo a mosaic instability. In the second mosaic instability that begins at 46\% overall DOD, a group of six particles undergo fast lithiation, and subsequently the cell DOD rises to 50\%; see Fig.\ \ref {insert}(iv). The successive mosaic instability involves a smaller number of particles undergoing fast lithiation because the number of particles at intermediate concentrations becomes smaller. 

\begin{figure}[htp] 
\begin{center}
\includegraphics[width=.75\textwidth]{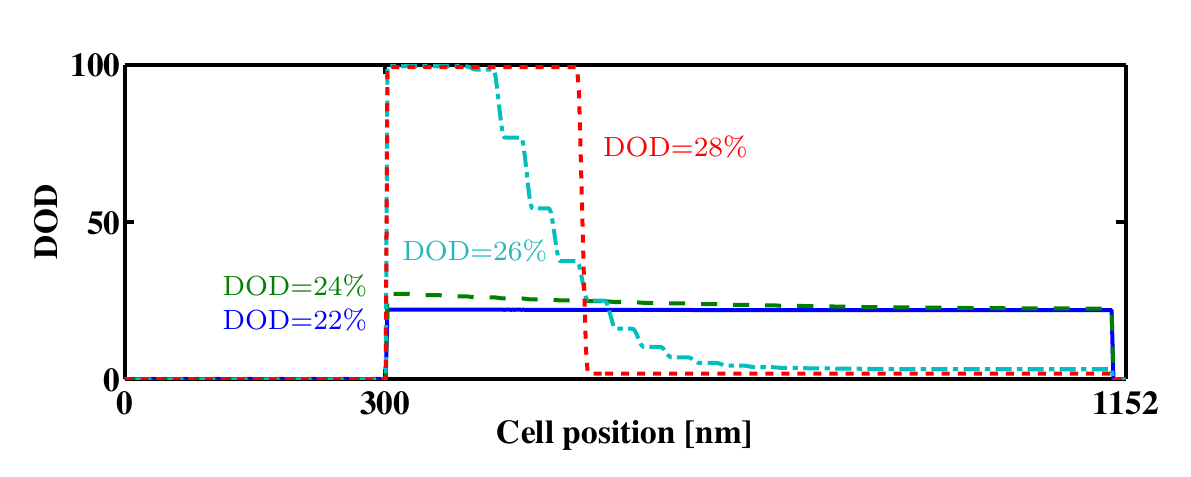}
\end{center}
\caption{DOD (equivalent to Li concentration) averaged over the particle region in a cross section in the x-y plane for the cell DOD in the range from 22\% (corresponding to Fig.\ \ref{insert}(b)(i)) to a DOD of 28\% (corresponding to Fig.\ \ref{insert}(b)(ii)) with a 2\% interval.}
\label{1mosaic}
\end{figure}

The process observed, which we here refer to as a group-by-group mosaic instability, is now explained. After some initial concentration accumulation, the first mosaic instability begins. As mentioned before, the particles closer to the separator obtain a slightly higher DOD. Because of the non-monotonic potential, the particles with a higher DOD within the spinodal region have a larger driving force to lithiate. Thus, a small difference in the DOD of the particles is amplified initiating fast lithiation. The process of the first mosaic instability is shown in Fig.\ \ref{1mosaic}, where we can see how a small initial difference in the DOD of the particles is amplified, triggering the instability. The voltage rises due to the increasing the equilibrium potential of the lithiating particles. Once ($\phi_s - \phi_l$), which is nearly uniform throughout the cell at low currents, is higher than the equilibrium potential of the particles that are not lithiating, the non-lithiating particles become thermodynamically driven toward delithiation. 
Therefore, this leads a group of particles to reach a nearly fully lithiated and another group a nearly fully delithiated state.  After the Li redistribution, the delithiated particles start lithiating again in a nearly uniform manner until the next mosaic instability is triggered, resulting in the intermittent group-by-group mosaic instability.

\begin{figure}[htp] 
\begin{center}
\includegraphics[width=.75\textwidth]{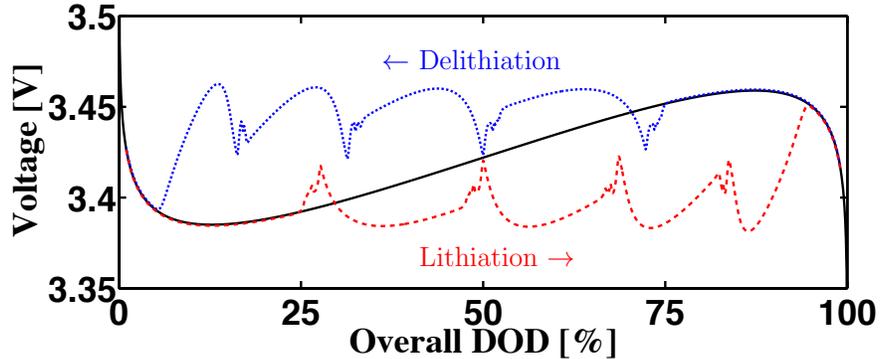}
\end{center}
\caption{Voltage measured during lithiation and delithiation when $i_0$ is assumed independent of the activity of the cathode. The dashed red and dotted blue lines represent the voltage measured during lithiation and delithiation, respectively. (Colors are only shown in the online version.)} 
\label{insertvolt}
\end{figure}

Figure \ref {insertvolt} shows the cell voltage profile during a lithiation-delithiation cycle, where the red and blue curves represent the lithiation and delithiation, respectively, and the black curve represents the system equilibrium potential. The two voltage curves are antisymmetric and the magnitude of the overpotential of the lithiation and the delithiation are the same when the DOD of the lithiation is equal to the state of charge ($\rm SOC = 1-DOD$) of the delithiation. Five sudden rises and drops of the voltage curve during both lithiation and delithiation are observed, which correspond to five discrete phase transformation instabilities. The voltage fluctuates around the lower spinodal point during lithiation. Conversely, it fluctuates around the higher spinodal point during delithiation. This leads to a voltage hysteresis between lithiation and delithiation.

\section {Results: lithiation and delithiation with activity-dependent exchange current density}
 \label {sec:var_j0}

Now, we include the dependence of the exchange current density on the activity of the particles as in Eq.~\eqref {i_0}. As will be shown later, this leads to significant asymmetric dynamics between lithiation and delithiation. In this set of simulations, we investigate how the applied current affects the particle interactions. Two different aspects are analyzed in the following subsections: (A) the lithiation and delithiation dynamics and their corresponding voltage fluctuations, and (B) the salt concentration and electrostatic potential in the electrolyte.
 
\subsection {Lithiation and delithiation dynamics and the corresponding voltage response}
\begin{figure}[htp] 
\begin{center}
\includegraphics[width=\textwidth]{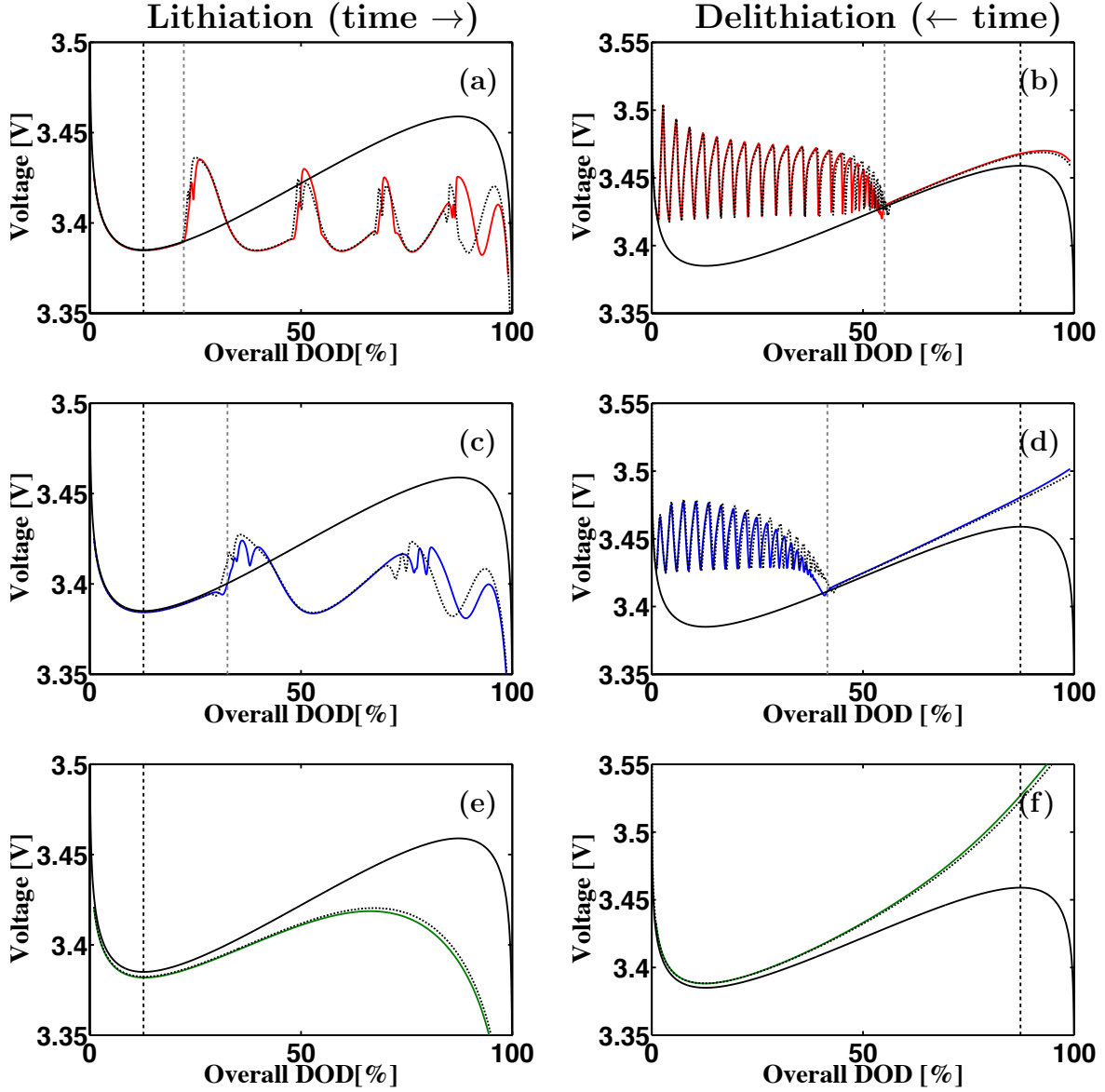}
\end{center}
\caption{Voltage vs.\ overall DOD during lithiation (left) and delithiation (right) at different currents. (a),(d) at $\overline i  = 2\% ~ i_0'$, (b),(e) at $\overline i  = 5\%~ i_0'$ and (c),(f) at $\overline i  = 20\% ~ i_0'$. The black curve corresponds to the equilibrium potential, the color curves to the cells voltage and the dashed lines to the corresponding simulations using a porous electrode model, which are discussed in Sec. \ref{PET}. The vertical black dashed lines indicate the lower and higher spinodal points during lithiation and delithiation, respectively. The vertical gray dashed lines the onset of the mosaic instability. (Colors are only shown in the online version)}
\label{volt_varj0}
\end{figure}

Here, we describe the lithiation and delithiation dynamics with an activity-dependent exchange current density in the Butler-Volmer equation. Note that in this section we only describe the dynamics observed, and the detailed analysis will be deferred to Sec.\ \ref{sec:analysis}. Figures \ref {volt_varj0}(a),(c),(e) present the voltage for lithiation and (b),(d),(f) for delithiation at different currents. 

The lithiation dynamics is similar to the case where the current density is independent of the activity, presented in Sec.\ \ref{sec:const_j0}.  In both cases, we observe group-by-group lithiation.  Figure \ref {volt_varj0}(a) shows the voltage curve for $\overline i  = 2\%~i_0'$ (C/12 rate), which is the same applied current as in the previous section.  The voltage exhibits five primary spikes during this process, indicating five sets of mosaic instabilities.  
As mentioned in Sec.~\ref{thermo}, purely based on the thermodynamics of the material, the phase separation is expected to occur at the spinodal point. However, significant deviations between the onset of the mosaic instability and the spinodal points are observed in our simulations. This deviation is referred to as the ``concentration overshoot'' hereafter. In Fig.~\ref{volt_varj0}(a), we observe that the mosaic instability occurs at 22\% DOD  (indicated by the vertical gray dashed lines), which is 9\% DOD higher than the lower spinodal point at $\sim$13\% DOD  (indicated by the black dashed vertical lines). Figure \ref {volt_varj0}(c) shows lithiation at $\overline i  = 5\%~i_0'$ (C/4.7 rate). Here, only three primary spikes in the voltage curve are present, indicating three sets of mosaic instabilities. From this observation, it can be deduced that each group contains more particles compared to the case where $\overline i  = 2\%~i_0'$. 
The concentration overshoot increases with an increasing magnitude of the current. In this case, the mosaic instability begins at 33\% DOD, which corresponds to a concentration overshoot of 20\% DOD. 
At a sufficiently high current, all the particles in the computational domain lithiate together and the interactions between the particles are suppressed. This is shown in Fig.\ \ref {volt_varj0}(e), which correspond to lithiation at $\overline i  = 20\%~i_0'$ (C/1.2 rate). In this case, there are no spikes in the voltage curve, showing no mosaic instability in the concentration evolution. 

\begin{figure}[htp] 
\begin{center}
\includegraphics[width=.75\textwidth]{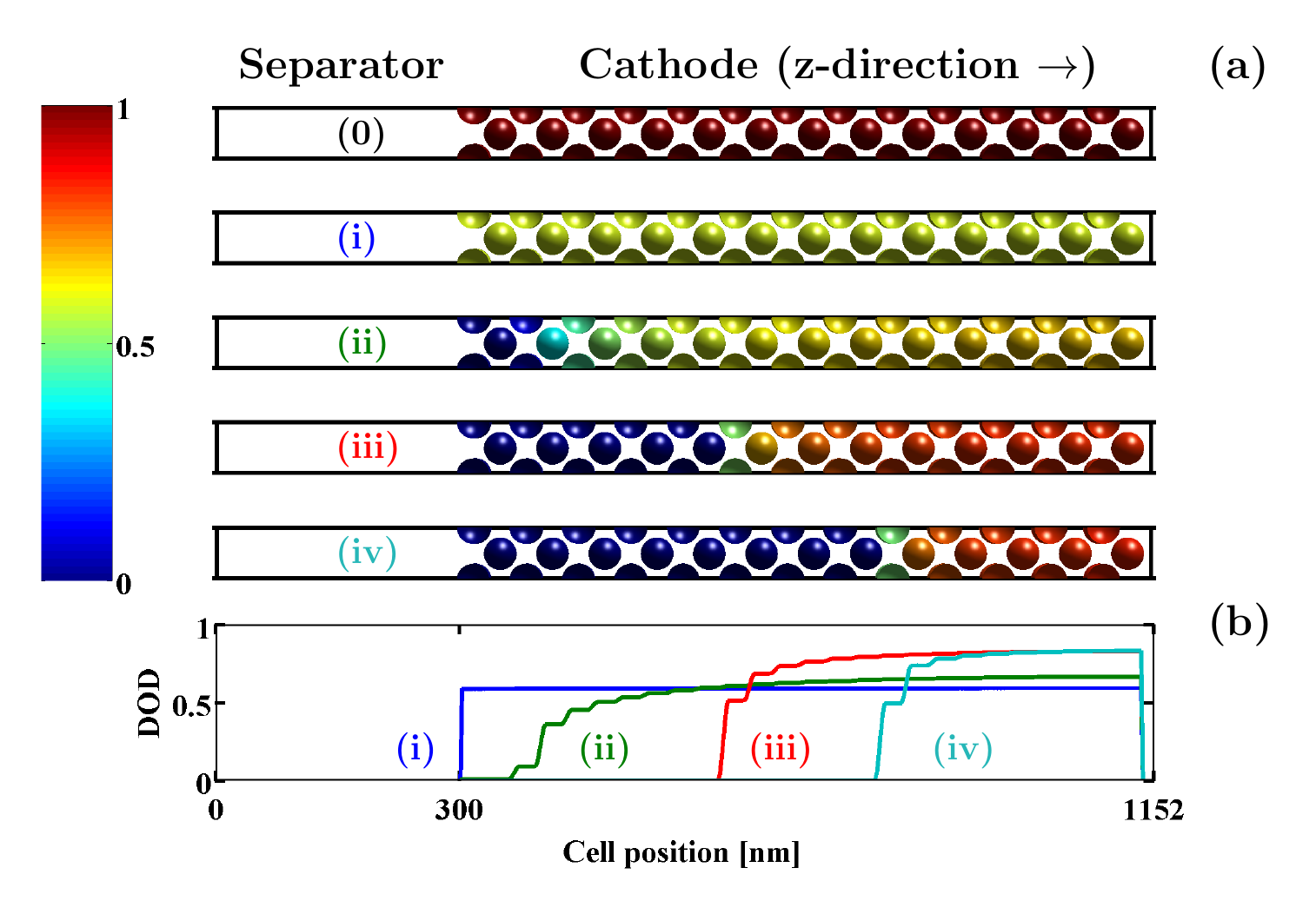}
\end{center}
\caption{Delithiation with exchange current density dependent of DOD. (a) Surface Li concentration of each individual particle layer. (0) Initial condition at $ {\rm DOD} = 98\%$. (i) All the particles delithiate in a fairly uniform manner ($ {\rm DOD} = 59\%$). (ii) The particles closer to the separator fully delithiate, releasing lithium that is absorbed by the remaining particles, some of which will return to a nearly fully lithiated state ($ {\rm DOD} = 54\%$). (iii) More particles become fully delithiated and the particles that were absorbing lithium return to a Li rich phase ($ {\rm DOD} = 48\%$). (iv) The particles continue to delithiate layer by layer ($ {\rm DOD} = 30\%$). As in Fig.~\ref{insert}, the surface concentration approximates the bulk concentration. (b) DOD averaged over the particle region in a cross section in the x-y plane.} 
\label{extract}
\end{figure}

On the contrary, the dynamics during delithiation differs substantially.  Figure \ref {extract} shows the snapshots of Li concentration evolution during delithiation at four different DODs at also $\overline i  = 2\%~i_0'$. The particles delithiate in a fairly even manner until reaching a DOD of around 59\%; see Fig.\ \ref {extract}(i). At this DOD, the first mosaic instability begins. At about 54\% DOD, the first three layers of particles are fully delithiated and the particles far away from the separator start absorbing lithium from those close to the separator; see Fig.\ \ref {extract}(ii). This is followed by particle-by-particle delithiation in the computational domain (which is equivalent to a layer-by-layer transformation because of the periodic boundary condition along the x- and y-directions), initiating from the separator-cathode boundary and moving toward the cathode current collector; see Fig.\ \ref {extract}(iii) at 48\% DOD.  Such a layer-by-layer phase-front movement continues until the entire cell is fully delithiated.  Figure \ref {extract}(iv) shows the concentration at 30\% DOD. The voltage curve for delithiation at this current is shown in Fig.\ \ref {volt_varj0}(b), where each spike corresponds to a fast delithiation event of one particle layer.
Here, the mosaic instability occurs at 55\%, which corresponds to a concentration overshoot of 32\% DOD. At  $\overline i  = 5\%~i_0'$, the instability occurs at DOD of 42\% and therefore the overshoot is 45\% DOD in magnitude ( = $\sim$87\% DOD at higher spinodal - 42\% DOD at first instability) ; see Fig.\ \ref {volt_varj0}(d). Unlike lithiation, the larger overshoot does not affect the number of particles undergoing fast lithiation because it still occurs as a layer-by-layer mosaic instability. At a sufficiently high current the mosaic instability is also suppressed; see Fig.\ \ref {volt_varj0}(f) at  $\overline i  = 20\%~i_0'$. 

\subsection{Salt concentration and electrostatic potential and in the electrolyte}
\begin{figure}[htp] 
\begin{center}
\includegraphics[width=\textwidth]{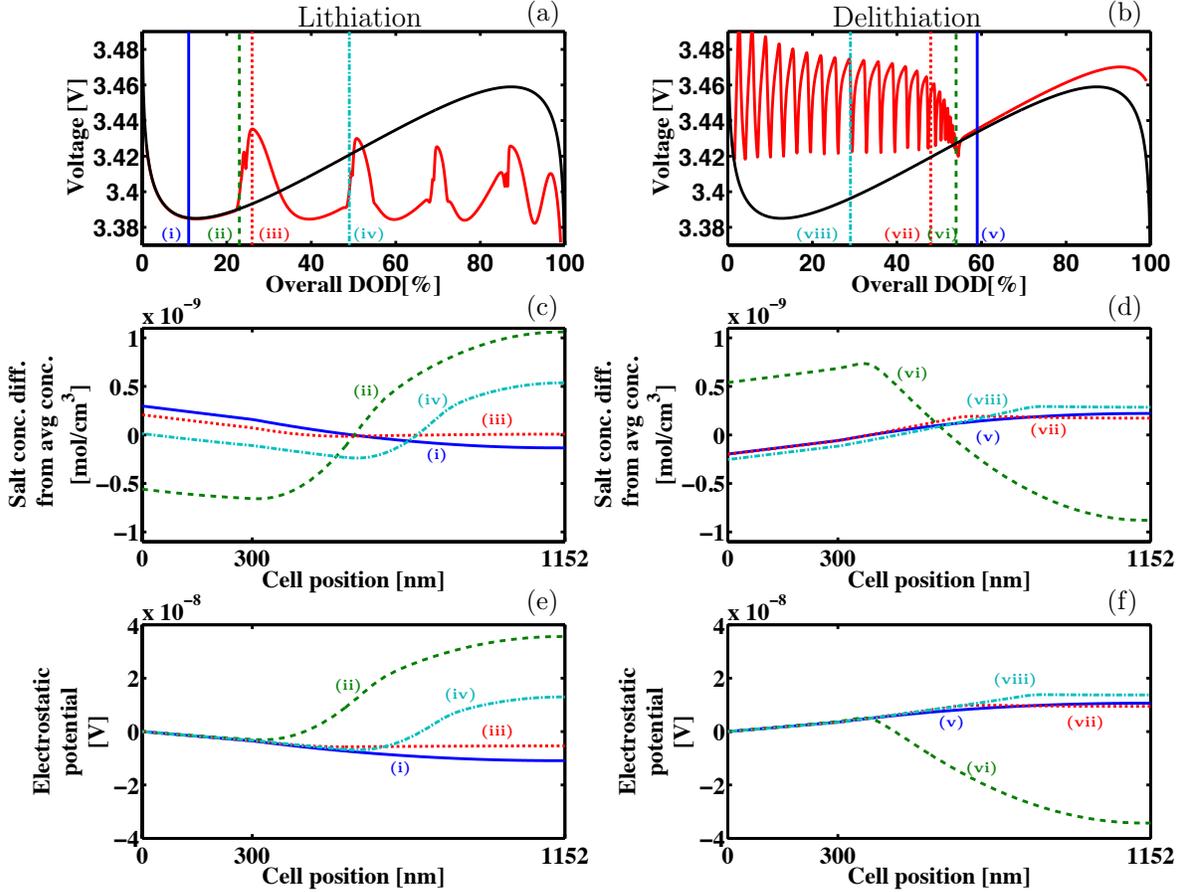}
\end{center}
\caption{ (a) and (b) Cell voltage during lithiation and delithiation, respectively at $\overline i  = 2\%~i_0'$. The vertical lines indicate the DODs at which salt concentration and electrostatic potential in the electrolyte are measured. (c) and (d) Salt concentration difference from the average concentration over the electrolyte region in a cross section in the x-y planes during lithiation and delithiation, respectively. (e) and (f) Electrostatic potential over the electrolyte region in a cross section in the x-y planes during lithiation and delithiation, respectively. The electrostatic potential at the anode is taken to be 0 V as a reference. For lithiation, (i) the solid blue line corresponds to 11\% DOD, (ii) the green dashed line to 23\% DOD, (iii) the red dotted line to 49\% DOD and (iv) the dashed-dotted cyan line to 68\% DOD. For delithiation, the same DODs as in Fig.\ \ref{extract} are used: (v) the blue solid line corresponds to 59\% DOD, (vi) the green dashed line to 54\% DOD, (vii) the red dotted line to 48\% DOD and (viii) the dashed-dotted cyan line to 30\% DOD. (Colors are only shown in the online version)}
\label{CLVL}
\end{figure}

Now, we analyze the effect of the mosaic instabilities on the salt concentration and the electrostatic potential of the electrolyte. For this purpose, we consider lithiation and delithiation at $\overline i  = 2\%~i_0'$. In Figs.\ \ref{CLVL}(a) and (b), the voltage for lithiation and delithiation, respectively, is shown again, along with markers noting the different states examined in this analysis.  Figures \ref{CLVL}(c) and (d) show the profiles of the salt concentration in the electrolyte at four different times during lithiation and delithiation, respectively. The curves are represented with the average concentration in the electrolyte on each slice of the x-y planes. 

During lithiation before the mosaic instability occurs, the salt concentration decreases gradually from the separator to the cathode current collector; see Fig.\ \ref{CLVL}(c) curve (i). When the first mosaic instability takes place, Li is absorbed rapidly on the particles near the separator, causing the salt concentration in the surrounding electrolyte to drop. Meanwhile, the particles closer to the cathode current collector eject Li into the electrolyte resulting in a rise in the salt concentration of that region. Consequently, a concentration increase from the cathode current collector side to the anode side is observed; see Fig.\ \ref{CLVL}(c) curve (ii).
After the first mosaic instability, the salt concentration returns to a similar state as prior to the instability; see Fig.\ \ref{CLVL}(c) curve (iii). In the second mosaic instability, the number of particles undergoing fast lithiation and the number of particles releasing Li is reduced. As a consequence, both the rise and drop in the salt concentration is reduced. Furthermore, the particles undergoing fast lithiation are located closer to the cathode current collector, causing the location of the concentration increase to also be shifted. In addition, the number of particles releasing Li is reduced. As a consequence, both the magnitude of the rise and drop in the salt concentration is reduced and occurs closer to the cathode current collector; see Fig.\ \ref{CLVL}(c) curve (iv). The subsequent instability is accompanied by smaller variations with the transition moving closer to the cathode current collector.

For the delithiation process, the salt concentration curve before the first mosaic instability is smooth but has a slope with an opposite sign from that of lithiation, since the flow of ions is in the opposite direction; see Fig.\ \ref{CLVL}(d) curve (v). In the first mosaic instability, the salt concentration of the electrolyte near the anode rises while the concentration near the cathode current collector decreases; see Fig.\ \ref{CLVL}(d) curve (vi). During the remainder of the delithiation, the concentration becomes nearly constant near the cathode current collector. This is because the Li concentration in the particles away from those that are reacting remain relatively unaffected during the subsequent mosaic instabilities. As previously described, the layer-by-layer phase-front moves continuously during delithiation, resulting in a continuous movement of the transition region of the concentration curve, where the slope changes rapidly; see Fig.\ \ref{CLVL}(d) curves (vii) and (viii).

We now describe the electrostatic potential in the electrolyte. Figure \ref{CLVL}(e) shows the electrostatic potential for lithiation. The same states (i)-(iv) as in the salt concentration are presented here.  The behavior of the electrostatic potential matches qualitatively with that from the salt concentration. The only difference is that all the curves align at the anode because the potential at the anode is set to 0 as the boundary condition. Similar explanations as for the salt concentration can be used to describe the behavior of the electrostatic potential.
The absorption of Li ions from the electrolyte into the particles causes not only a decrease in the salt concentration, but also a decrease in the electrostatic potential. As observed in Eq.\ \eqref{curr_cons_l}, this relation is maintained away from the interfacial regions as long as $(D_+-D_-)$ is negative. During the mosaic instability, some of the particles extract lithium into the electrolyte, locally increasing the electrostatic potential in a similar way as observed with the salt concentration.
Figure \ref{CLVL}(f) presents the electrostatic potential during delithiation also at the same times (v)-(viii) as for the salt concentration. An analogous explanation applies for delithiation. 

\section{Analysis}
\label {sec:analysis}

Here, we analyze and further discuss the results presented in the prior sections, primarily focusing on cases where the exchange current density is dependent on the cathode activity, unless otherwise noted. Three different aspects are included in this analysis: (A) the origin of the concentration overshoot, (B) the determinant of the group size in lithiation and (C) the origin of the asymmetry in lithiation and delithiation behavior.

\subsection{Concentration overshoot}
\label{CCD}

\begin{figure}[htp] 
\begin{center}
\includegraphics[width=1\textwidth]{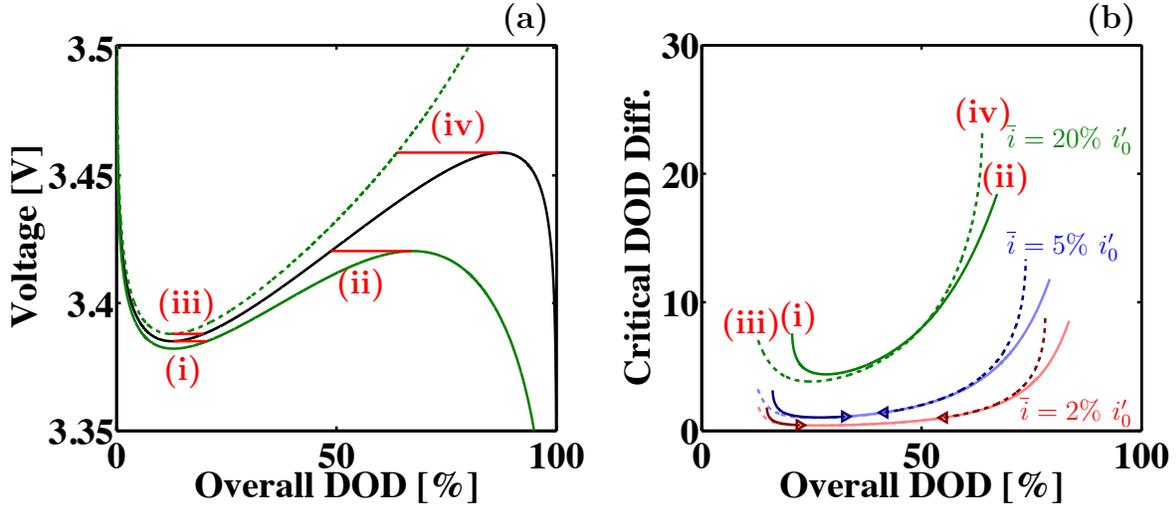}
\end{center}
\caption{(a) Applied voltage for lithiation and delithiation at $\overline i  = 20\%~i_0'$ represented by the solid green line and the dashed green line, respectively. The minimum DOD difference required for the mosaic instability to begin (i)-(ii) during lithiation, and (iii)-(iv) during delithiation is here indicated. The black line represents the equilibrium potential. (b) Minimum DOD difference as a function of the overall DOD for $\overline i  = 2\%~i_0'$, $\overline i  = 5\%~i_0'$ and $\overline i  = 20\%~i_0'$, represented here by a red, blue and green line, respectively. Right-pointing and left-pointing triangles represent the first mosaic instability during lithiation and delithiation, respectively. The darker lines represent the DODs before the instability occurs, while the lighter lines indicate the regions after it occurs. The solid line represents lithiation and the dashed line delithiation. (Colors are only shown in the online version)}
\label{critical}
\end{figure}

As described in Sec \ref{sec:var_j0}, a larger concentration overshoot occurs during delithiation in comparison to lithiation, and in both cases the overshoot increases with current. Now, we perform a simplified analysis to illustrate the origin of this change.  For this analysis, we assume a cell in which a group of particles react uniformly and one adjacent particle has a different concentration from the rest. We will denote the particle as ``particle A.'' In this hypothetical cell, there is a significant amount of particles such that the DOD of particle A does not affect the DOD of the cell. In order for the group of particles with uniform Li concentration to lithiate at a given current, an applied voltage, ``$V_x$,'' is required. At the same time, the adjacent particle must have an equilibrium potential lower than $V_x$ to have a driving force for delithiation while the other particles are lithiating. The Li concentration of the particle at which this occurs can be determined by the intersection of the equilibrium potential curve and a horizontal line of value $V_x$. This construction is shown in Fig.~\ref{critical}(a) for (de)lithiation at $\overline i = 20\%~i_0'$. The curves for lithiation and delithiation are different because of the activity dependent $i_0$ (Fig.\ \ref{i0}).

We now describe the composition difference required for the initiation of the mosaic instability during lithiation and delithiation. As previously mentioned, in order for particle A to delithiate while the group of particles lithiates, particle A must have an equilibrium potential lower than $V_x$. Thus, the lengths of line (i) and line (iv) in Fig.~\ref{critical}(a) represent the composition difference required at the first DOD at which the instability is possible for lithiation and delithiation, respectively. For lithiation, the concentration of the cell is indicated by the right end of line (i) (21\% DOD) while the corresponding concentration of the particle A must be at the left end of line (i) (13\% DOD). Therefore a composition difference of 8\% DOD is required in order for the instability to take place. The mosaic instability could not occur at a cell DOD smaller than 21\% since the applied voltage below that composition is outside the voltage window of the equilibrium potential.  
For delithiation, the initial concentration of the cell is indicated by the left end of the line (iv) (64\% DOD) and the concentration of the lithiating particle by the right end of the line (41\% DOD), therefore a composition difference of 23\% DOD is required for the instability to begin.  By comparing lithiation to delithiation, the following is observed: First, the DOD at which the mosaic instability can occur is farther away from the spinodal points during delithiation compared to lithiation. Second, once the instability can occur, it also requires a larger concentration difference in delithiation compared to lithiation. These two reasons make the concentration overshoot in delithiation larger. 

Now, we compare the critical concentration difference at different currents. Figure \ref{critical}(b) shows the critical composition difference as a function of the cell DOD for the three different applied currents for lithiation and delithiation. The curves are calculated based on the difference between the equilibrium potential and the theoretical (de)lithiation voltage of the particles in the absence of mosaic instability, as described above. Again, the difference between lithiation and delithiation originates from the activity dependence of $i_0$. The darker curves indicate the range of the cell DODs before the onset of the mosaic instability relevant for the simulations presented earlier, while the lighter lines indicate the theoretical calculations beyond the onset of the mosaic instability. The right- and left-pointing triangular markers on the blue and red lines indicate the points of onset of the first mosaic instability in lithiation and delithiation, respectively. The required composition difference increases as the loading current increases, and thus the composition at which instability can first occur deviates farther from the spinodal points. As a consequence, a larger concentration overshoot occurs with a higher current. Note that, even though the mosaic instability could occur at $\overline i = 20\%~i_0'$, it does not occur if the required composition difference is not reached (as is the case in our simulation). Thus, the analysis provides only the minimum overshoot required, not the actual overshoot. If the applied current is high enough such that the cell voltage is outside the voltage window (where the mosaic instability, {\it i.e.}, full and empty particles, is favored), the mosaic instability will be suppressed, independent of the composition differences. 
 
\subsection{The determinant of the group size in lithiation}  
 \label {groupsize} 
Having established qualitatively the relationship between the concentration overshoot and the current, we now present an analysis to determine the fraction of particles reacting during each mosaic instability for a given concentration overshoot. Note that this analysis is rate independent and in here the transport of the salt in the electrolyte is assumed not to be a limitation. As previously described, during the mosaic instability upon lithiation, the particles with a Li concentration in the spinodal region (the ``active particles") exchange Li with each other. By such exchange process, the particles reach either an almost fully lithiated or an almost fully delithiated state. At the onset of the mosaic instability, the particles outside the spinodal region (the ``inactive particles") are only those that have already undergone fast lithiation. Thus, we assume that the inactive particles have a DOD of 100\%. Given the DOD of the cell, $DOD_{cell}$, the average DOD of the active particles, $DOD_{active}$, can be approximated from the expression
\begin {equation}
\label {size1}
\frac {N_{inactive}} {N_{tot}} 100\% + \frac {N_{active}}{N_{tot}} DOD_{active}   = DOD_{cell},
\end {equation}
where $N_{inactive}$, $N_{active}$ and $N_{tot}$ are the number of inactive, active and total particles, respectively. Next, we choose $DOD_{cell}$ at which we can approximate the DOD of the lithiating and delithiating particles. This point corresponds to the DOD at which the voltage peaks occur. If several small peaks occur as part of one main instability event, we take the latest one, as this is the point when concentration can be best approximated. As illustrated in Sec.\ \ref{sec:2part} (see Figs.\ \ref{insert2p}(b) and (c) point (iv)), at this DOD the lithiating particles are close to the higher spinodal point ($\sim$87\% DOD). The actual location of the peak depends on the magnitude of the applied current and the number of particles reacting. Due to the activity dependence of $i_0$, the voltage peak shifts to lower DODs at higher currents. However, for simplicity in the calculation, we ignore this shift. The delithiating particles are at a low concentration that we refer as $DOD_{LC}$ ($\sim$0\% at the DOD of Fig.\ \ref{insert2p}(b) point (iv)). The concentration of the delithitated particles depends on the height of the voltage peaks, as their equilibrium potential has to remain lower than the applied voltage. $DOD_{LC}$ is here approximated as the DOD at which the equilibrium potential below the spinodal point is equal to the applied voltage.

The fraction of active particles that lithiate simultaneously in a given group can be therefore approximated as
\begin {equation}
\label {size2}
\frac {N_{group}} {N_{active}} = \frac {DOD_{active}-DOD_{LC} } {DOD_{HS}-DOD_{LC}},
\end {equation}
where $N_{group}$ is the number of particles in the group, and $DOD_{HS}$ the DOD of the higher mosaic instability. Substituting Eq.\ \eqref{size2} into Eq.\ \eqref{size1}, the fraction of particles that react simultaneously can be expressed as
\begin {equation}
\frac {N_{group}} {N_{active}}  =  \frac {N_{tot}DOD_{cell} -  N_{inactive} - N_{active}DOD_{LC} } {N_{active} (DOD_{HS}-DOD_{LC})}.  
\end {equation}
The predictions from this analysis and from our simulations with an activity-dependent $i_0$ are provided in Table \ref {prediction} for comparison. Note that this analysis could also be used for lithiation with an activity-independent $i_0$, and a similar analysis could be performed for delithiation with an activity-independent $i_0$. In the earlier groups (1$^{\rm st}$-3$^{\rm rd}$ in the case of $i=2\%~i_0' $ and 1$^{\rm st}$-2$^{\rm nd}$ in the case of $i=5\%~i_0' $ ), the estimated fractions are in good agreement with the simulations. However, in the later groups, the estimates are less accurate. There are two primary reasons for this disagreement. (1) The small number of remaining active particles limits the results of the fractions in simulations, for example there are only two particles in the 5$^{\rm th}$ group of $i=2\%~i_0' $ and therefore the resulting fraction can only take the value of either 50\% or 100\%. (2) Because of the smaller number of particles reacting in the later groups, those particles undergo fast lithiation at a higher rate, which makes the shift of the voltage peak larger. For example, the DOD at which the voltage peaks in the case of $i=2\%~i_0' $ for the first group is $\sim$84\% while for the last group is $\sim$64\%. With the consideration of this deviation, the calculated fraction of particles reacting in the last group increases from 63\% to 87\%, which is closer to the fraction observed in the simulation.

\begin{table}[htp]
\begin{tabular}{|c|c|c|c|c|} 
\hline
Applied current & Group number & Cell DOD & Predicted fraction & Fraction in simulations \\
\hline
& 1$^{\rm st}$ & 26\% & 29\% & 31\% (8/26)\\
& 2$^{\rm nd}$  & 51\% & 32\% & 33\% (6/18)\\
$\overline i=2\%~i_0' $  & 3$^{\rm rd}$ & 70\% & 39\% & 41\% (5/12) \\
& 4$^{\rm th}$  & 87\% & 59\% & 71\% (5/7)\\
& 5$^{\rm th}$ & 97\% & 63\% & 100\% (2/2)\\
 \hline
& 1$^{\rm st}$ & 40\% & 45\% & 46\% (12/26)\\
$\overline i=5\%~i_0' $ & 2$^{\rm nd}$  & 81\% & 73\% & 78\% (11/14)\\
& 3$^{\rm rd}$ & 94\% & 57\% & 100\% (3/3)\\
 \hline
\end{tabular}
\caption {Fraction of the active particles that react simultaneously obtained by the calculation presented here, compared to the fraction observed in the simulations.}
\label{prediction}
\end{table}
   
\subsection{Origin of the asymmetry in lithiation and delithiation behavior}
 \label {sec:origin} 
 In this section, we elucidate the origin of the asymmetry of the mosaic instability during lithiation and delithiation when the current is sufficiently low, observed in Sec.\ \ref{sec:var_j0}. This asymmetry is a manifestation of a competition between two processes that occur simultaneously: (1) an intermittent Li redistribution among the particles and (2) a constant (de)lithiation of the cell due to the externally applied voltage (which is varied to maintain the desired current). Which of these processes becomes dominant depends on the value of $i_0$, the exchange current. 

\begin{figure}[htp] 
\begin{center}
\includegraphics[width=1\textwidth]{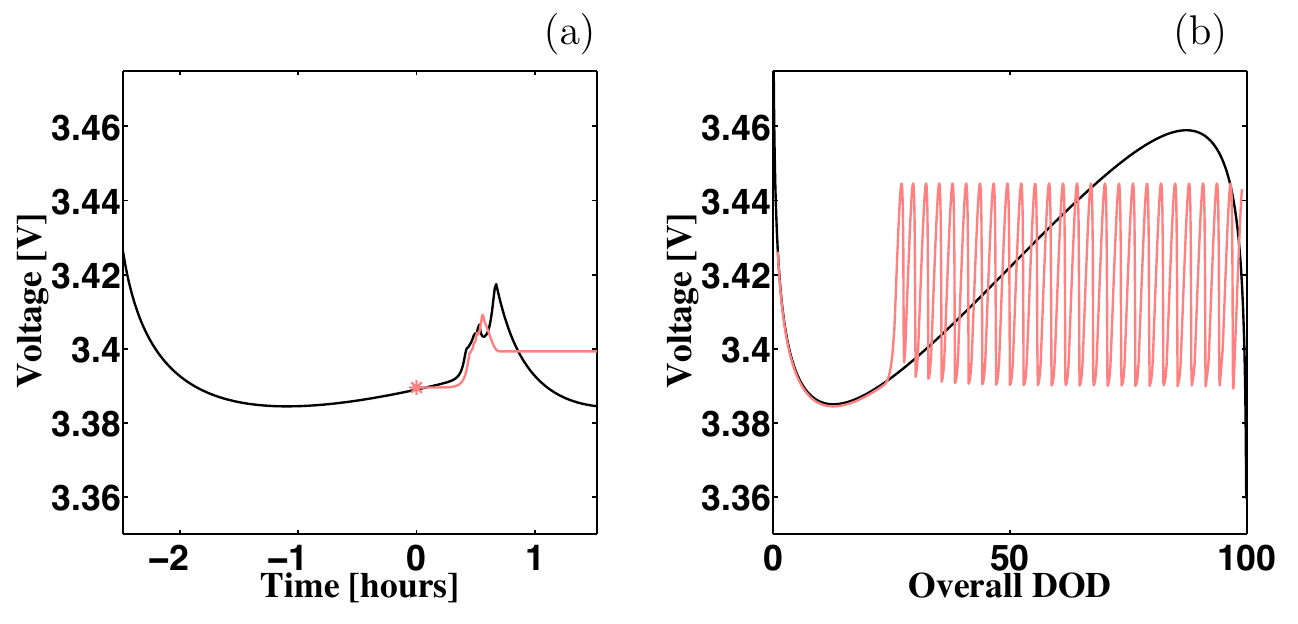} 
\end{center}
\caption {(a) Relaxation of a cell with a DOD of 22\%. The light red curve represents the voltage of the relaxed cell. The black curve indicates the voltage of the lithiating cell at $\overline i=2\%~i_0'$ where $i_0$ is activity independent (corresponding to the data shown in Fig.\ \ref{insertvolt} in the range from 1\% to 35\% DOD). The light red asterisk indicates the beginning of the relaxation of the cell. (b) Lithiation of the cell at $\overline i=2\%~i_0'$ where $i_0$ is activity independent and all the backflow fluxes from the particles to the electrolyte are hindered. The light red curve indicates the voltage of the cell and the black curve the equilibrium potential. (Colors are only shown in the online version)}
\label{analysis}
\end{figure}

To illustrate these two processes and the resulting dynamics, we conduct simulations including only one of the processes. The insight gained is employed to facilitate our understanding of the origin of the asymmetric dynamics.  First, to analyze Li redistribution among the particles without an applied current, we conduct a simulation in which a cell is relaxed with a nearly uniform DOD of 22\% (taken from the partially lithiated cell with an activity independent $i_0$ and $\overline i=2\%~i_0'$, presented in Sec.\ \ref{sec:const_j0}; see Fig.\ \ref{insert}(a)(i)). During the process of relaxation, a cell voltage is imposed to maintain a zero net current through the current collectors. Shown in Fig.\ \ref{analysis}(a) is the cell voltage during the relaxation. Throughout the relaxation where five particles reach a lithiated state and 21 particles a delithiated state, one primary sudden rise of the voltage occurs as a consequence of changing the Li concentration of the particles and their corresponding equilibrium potential. This shows a natural tendency for the system to transit from an activated unstable state with partially lithiated particles to a coexistence of lithium rich and lithium poor particles. Note that the shape of the voltage rise and drop observed in relaxation is very similar to those observed during lithiation at $\overline i=2\%~i_0'$ with an activity independent $i_0$. 

Next, we focus on the effect of the applied current by prohibiting an opposing reaction ({\it i.e.}, Li back-flow into the electrolyte during lithiation of the cell).
\begin{equation}
r{Li} = 
\begin{cases}
r_{Li}^{BV} &~~{\rm if}~~ r_{Li} \ge 0 \\
0 & ~~{\rm if}~~ r_{Li} < 0.
\end{cases}
\end{equation}
The simulation is conducted at $\overline i=2\%~i_0' $, and $i_0$ is activity independent. The resulting voltage response indicates that the rapid lithiation occurs one particle at a time. We observe 26 sudden rises (and drops) of the voltage, corresponding to the fast lithiation of the 26 particles; see Fig.\ \ref{analysis}(b).
The process begins in the same manner as the case when the back-flow is allowed (described in Sec.\ \ref{sec:const_j0}). The lithiation increases the DOD of the particles nearly uniformly. Upon reaching the concentration overshoot to trigger the first mosaic instability, the first particle begins lithiating rapidly. 
However, because delithiation is prohibited here, the current is a result of lithiation reaction only, which needs to be slower than the case where delithiation accompanies the process. Thus, a higher voltage, which reduces the lithiation reaction rate, is observed in Fig.\ \ref{analysis}(b).  The peaks in this figure are higher than those observed in Fig.\ \ref{insertvolt}.  This larger voltage during fast lithiation leads to a more rapid amplification of the concentration gradient and particle-to-particle variation of the driving force for lithiation.  This amplification causes the ``runaway" reaction of a particle and consequently a particle-by-particle mosaic instability.  The lithiation of the subsequent particle can only occur when the rapidly lithiating particle is nearly fully lithiated and the cell voltage decreases to a value below  the equilibrium potential of the next particle.  This process is repeated and, as a result, the lithiation proceeds in a layer-by-layer manner.

With the background provided by the discussion of the two processes above, we elucidate mechanisms that lead to the observed asymmetric dynamics in our simulations for the case with the activity-dependent $i_0$. During lithiation, the particles undergo mosaic instability in a group-by-group manner, similarly to the case of an activity-independent $i_0$ presented in Sec. \ref{sec:const_j0}.
The dynamics of this instability strongly resembles that of relaxation, showing that Li redistribution is significant in this case. During the mosaic instability, the delithiation of the active particles is strongly facilitated by a ``sufficiently large'' value of $i_0$ for the particle DOD in the range between an almost fully delithiated state and the onset of the mosaic instability. Here, we denote that $i_0$ is sufficiently large when the value for the particles toward delithiation is similar to or larger than that toward lithiation, at the onset of the instability. In other words, $i_0$ must be large enough to facilitate delithiation during the lithiation process. Note that this condition for $i_0$ is met for the lithiation simulations in the previous sections with both activity-dependent and activity-independent exchange current density, as well as for delithiation with activity-independent $i_0$; see Fig.\ \ref{i0}. In those cases, Li redistribution readily occurs and is the dominant mechanism for the dynamics of the mosaic instability.

In contrast, the delithiation behavior for an activity-dependent $i_0$, which shows layer-by-layer dynamics, resembles the case of prohibited Li redistribution. At the onset of mosaic instability upon delithiation, because the tendency for a particle to lithiate when a neighboring particle delithiates is week due to the value of $i_0$ (see Fig.\ \ref{i0}), the Li redistribution is hindered. As a result, when a lithiated particle undergoes fast delithiation and causes a voltage drop, other lithiated particles remain at a fairly constant concentration close to the higher spinodal point, without returning to the nearly fully lithiated state. The only exceptions are the initial instabilities in which the concentration overshoot is larger and therefore some redistribution occurs. The redistribution is kinetically hindered by the exchange current density at a concentration close to the higher spinodal point. Thus, the fast delithiation of a particle corresponds mostly to the constant extraction of Li out of the cathode without intermittent exchange of Li between the particles. This results in a layer-by-layer mosaic instability, similar to the simplified case where no opposing reaction was allowed.

In summary, our analysis indicate that, because of the asymmetric function of exchange current density, Li redistribution is facilitated upon lithiation of the cell, while redistribution is limited upon delithiation. This difference manifests as a group-by-group instability during lithiation, a thermodynamically favored behavior, and a layer-by-layer instability during delithiation, a kinetically controlled behavior. Note that, besides an asymmetric $i_0$, other factors such as an asymmetric equilibrium potential as the one presented by Malik {\it et al.} \cite{Malik:2011xd} or a transfer coefficient with a value different from 0.5 also lead to asymmetric dynamics \cite{Zeng:2013qy}.

\section{Comparison to the porous electrode model}
\label {PET}

In this section, we compare our results with those obtained by Ferguson and Bazant using a porous electrode model \cite{Ferguson:2012vk}. For this comparison, the simulation parameters in the porous electrode model were set effectively equal to those in our particle-level simulation and the equations described below were used. 

The porous electrode model used is in the pseudocapacitor limit.  That is, transport in the solid is fast compared to surface reactions and transport in the electrolyte.  For nanoparticles, this approximation is reasonable.  This allows the concentration profiles inside the particles to be neglected and the particles can be treated as sink terms.  Since the model averages over the volume of the electrode, what is referred to as a particle is actually a representative particle for that volume of the electrode, and all solid particles inside that volume are assumed to behave the same.  The accumulative Li concentration, $\overline {C_s}$, can be determined by the reaction rate:
\begin {equation}
\label {PET1}
\frac {\partial \overline {C_s}} {\partial t}  = a_p  {r_{Li}}. 
\end {equation}
where $a_p$ is the area to volume ratio of the particles. This equation governs the lithiation of the particles and substitutes Eq.\ \eqref{SimpleDiff} and  \eqref{SimpleDiff2} of our model. Here, 26 of these ``particles'' ({\it i.e.}, volumes of particles) are considered within the cathode. 

The average porosity of the electrode, $\epsilon$, defined as volume fraction of electrolyte with respect to the total volume of the cathode, is used to obtain the effective diffusivity and conductivity of the porous media via the Bruggeman empirical relation. In 1D, the salt concentration evolution in the electrolyte is described by
\begin {equation}
\label {PET2}
\epsilon \frac {\partial C_l} {\partial t} =  \frac {\partial} {\partial z} \bigg (  \epsilon D_{amb} \frac {\partial C_l} {\partial z} \bigg ) - (1-t_+) a_p r_{Li}. 
\end {equation}
Two primary differences between this equation and the original equation (Eq.\ \eqref{Cs_time3}) can be noticed. First, instead of spatially resolving the electrolyte that fills the electrode, the salt concentration is averaged over the volume. Second, the particle surfaces are no longer explicitly defined, and are replaced by a given value of particle surface area. 

In order to express the current density, we also need to account for the porosity, and therefore the original equation (Eq.\ \eqref {curr}) is modified to include this factor,
\begin {equation}
\label {PET3}
i  = -z_+ \upsilon_+ F \bigg[ \frac {F} {RT} (z_+D_+ - z_-D_-) \epsilon C_l \frac {\partial \phi_l } {\partial z} + (D_+ - D_- ) \epsilon \frac {\partial C_l } {\partial z}   \bigg].
\end {equation} 
Lastly, the current continuity equation (Eqs.\ \eqref {curr_cons_l}-\eqref {curr_cons_l2}) becomes
\begin {equation}
 \frac {\partial i} {\partial z}  = r_{Li}F a_p. 
 \end {equation} 
Here, it is assumed that the reaction occurs throughout the entire porous electrode. Detailed explanations of this model can be found in Ferguson and Bazant's work \cite{Ferguson:2012vk}.

The porous electrode models are computationally efficient and often capture the dynamical nature of the charge and discharge process within a relatively simple description. However, the simplification leads to a disadvantage that they do not directly allow investigation of microstructural details and resulting effects since they only consider average properties. Thus they require validation and examination of the limit of applicability. Figures \ref {volt_varj0}(a)-(f) show the comparisons between the porous electrode simulation (dashed lines) and the SBM simulations (solid lines). The two results are in remarkable agreement. They both capture the mosaic instabilities observed in the process, have a similar concentration overshoot and predict different dynamics between lithiation a delithiation. However, there is a small difference in the magnitude of the overpotential between the two simulations. This difference leads to a disagreement between the later mosaic instabilities during lithiation, and between the onsets of the first instabilities during delithiation. These differences can be attributed to various approximations involved in each method. On one hand, an artificial finite thickness is assigned to the particle-electrolyte interface in the SBM, and on the other hand, several simplifications are taken in the porous electrode model as described above. Both methods carry some degrees of small errors that, at the end, leads to the disagreement. For the error analysis in the SBM, one can find the information in the work of Yu {\it et al.} \cite{Yu:2012oq}.

The above analysis demonstrates that the models from the two different length scales accurately describe the same physical phenomenon of mosaic instability. Note that the agreement between the two models is partly due to the simplicity of the microstructure used here. These two models compliment each other: The particle-level model allows us to study more detailed electrochemical dynamics accounting for the complexity of microstructures \cite{Yu:fk}, which would not be revealed in a homogeneous porous electrode model. The porous electrode model allows us to study much larger cells, such as those from a commercial battery, which is not currently feasible with particle-level simulations.

\section {Conclusion}
 
In this paper, we have investigated the behavior of an array of single-sized particles that are not allowed to generate a phase boundary within a particle, despite the bulk thermodynamic driving force to do so.  Mosaic instabilities are observed when the current is sufficiently low.  Through analysis, the concentration overshoot was explained, and the group sizes of the mosaic instability were predicted.  Further careful examination elucidated the competition of two mechanisms: thermodynamic relaxation that leads to Li redistribution and to group-by-group phase transformation, and kinetically induced layer-by-layer phase transformation.  The asymmetry between lithiation and delithiation is attributed to the exchange current density model, appearing in the modified Butler-Volmer equation.  We also compared our simulation with the porous electrode model of Ferguson and Bazant \cite{Ferguson:2012vk}, which showed excellent agreement and provided further insights into the mechanism underlying the lithiation/delithiation dynamics, resulting from the model.

In normal battery cells, material and structural non-uniformities, such as defects in particles, distribution of particle sizes, electronic conductivity between particles and variation of salt concentration in the electrolyte, are present. As such, the mosaic instability becomes a local phenomenon instead of a cell-wide one, and is thus difficult to directly observe in experiments. In addition, the voltage response to an individual mosaic instability event could be washed out when averaged over the cell. Nonetheless, our work sheds insights into the dynamics of lithiation/delithiation at the particle level, which affects the macroscopic behavior of nanoparticulate phase separating cathodes. 

\vskip 0.1in
The authors thank the National Science Foundation for financial support under Contract No.\ DMS-0842504. Additionally, TRF and MZB thank the National Science Foundation for support under Contract No.\ DMS-0948071 and partial support of Samsung via the Samsung-MIT Program for Materials Design and Energy Applications. The computational resources were provided by the Extreme Science and Engineering Discovery Environment (XSEDE), which is supported by National Science Foundation grant number OCI-1053575, under allocation No.\ TG-DMR110007.

\end{document}